\documentclass[twocolumn,twoside,slac]{revtex4}
\usepackage{graphicx}
\usepackage{fancyhdr}
\begin{document}

\title{The CMS Integration Grid Testbed}

%

\author{Gregory E. Graham, M. Anzar Afaq, Shafqat Aziz, L.A.T. Bauerdick, Michael Ernst, Joseph Kaiser, Natalia Ratnikova, Hans Wenzel, Yujun Wu}
\affiliation{Fermi National Accelerator Laboratory, Batavia, IL, 60510-0500}
\author{Erik Aslakson, Julian Bunn, Saima Iqbal, Iosif Legrand, Harvey Newman, Suresh Singh, Conrad Steenberg}
\affiliation{California Institute of Technology, Pasadena, CA, 91125}
\author{James Branson, Ian Fisk, James Letts}
\affiliation{University of California, San Diego, CA, 92093}
\author{Adam Arbree, Paul Avery, Dimitri Bourilkov, Richard Cavanaugh, 
Jorge Rodriguez, Suchindra Kategari}
\affiliation{University of Florida, Gainesville, FL, 32601}
\author{Peter Couvares, Alan DeSmet, Miron Livny, Alain Roy, 
Todd Tannenbaum}
\affiliation{University of Wisconsin, Madison, WI, 57303}

\begin{abstract}
The CMS Integration Grid Testbed (IGT) comprises USCMS Tier-1 and
Tier-2 hardware at the following sites: the California Institute of Technology, 
Fermi National Accelerator Laboratory, the University of California at San 
Diego, and the University of Florida at Gainesville.  The IGT runs jobs
using the Globus Toolkit with a DAGMan and Condor-G front end.  The 
virtual organization (VO) 
is managed using VO management scripts from the European Data Grid (EDG).  
Gridwide monitoring is accomplished using local tools such as Ganglia 
interfaced into the Globus Metadata Directory Service (MDS)
 and the agent based 
Mona Lisa.  Domain specific software is packaged and installed using the Distribution After Release (DAR) tool of CMS, while middleware under the auspices 
of the Virtual Data Toolkit (VDT) is distributed using Pacman. During a continuous 
two month span in Fall of 2002, over 1 million official CMS GEANT 
based Monte Carlo events were generated and returned to CERN for analysis 
while being demonstrated at SC2002.  In this paper, we describe the 
process that led to one of the world's first continuously available, 
functioning grids.  
\end{abstract}

\maketitle

\thispagestyle{fancy}


\section{Introduction}

  The CMS Integration Grid Testbed (IGT) was commissioned by USCMS in 
the Fall of 2002 in order to provide a stable platform for 
integration testing of Grid middleware with the existing CMS 
Monte Carlo production environment.  CMS (Compact Muon Solenoid) is a
high energy physics detector planned for the Large Hadron Collider 
(LHC) at CERN, just outside of Geneva, Switzerland.  While CMS will not begin
taking data until after 2007, hundreds of physicists around the world
are taking part in Monte Carlo simulation studies of the detector and 
its potential for discovering physics.

The IGT is intended to comprise a small number of nodes at USCMS Tier-1 
and Tier-2 centers from hardware that is assigned to USCMS Facilities.  
Within the IGT, focused development 
and integration prototyping takes place in support of a set of middleware 
products and CMS applications to be used in a production grid setting.
Currently, the IGT actually comprises most of the available USCMS Tier-1 and 
Tier-2 resources.  This is because 
\begin{itemize}
\item This was the first time we have tried seriously to put Grid middleware
into production on this scale in a continuously available fashion.  
\item There was as yet no documentation or level of support in place 
appropriate to a production setting
\end{itemize}

The goals of this first running of the IGT are to prepare for a 
turnover of newly integrated software onto the first Production Grid
and to start writing documentation appropriate to a production setting,
while participating in official CMS Monte Carlo production.
The turnover is scheduled to happen in Spring 2003.  At that time, 
the resources assigned to the IGT will mostly migrate to the production grid (PG).\footnote{From time to time, it will happen that large amounts of resources will 
be temporarily assigned to the IGT for scalability tests in the integration
setting.} In addition, CMS has a development grid testbed (DGT) in order to 
provide a less stringent environment for more speculative development and
test deployments. With these three grids in place, USCMS is well positioned to deploy a
Grid with a view to continuous upgrades and improved software quality control
while maintaining a development environment conducive to new concepts and
experiments in Grid middleware.   This also provides an effective response to 
the declared intention of the LHC Computing Grid (LCG) to produce a continuously 
available 24x7 production grid for the LHC experiments.

\section{Hardware Included in the IGT}

Table \ref{tb:hardware} shows the available hardware on the IGT.  Participating sites
included the California Institute of Technology, Fermi National Accelerator Laboratory, 
University of California at San Diego, and the University of Florida at Gainesville as of
November 1, 2002.  A group from the LCG at CERN also joined the effort after November 15, 
2002.  While the University of Wisconsin at Madison Computer Science department did not
participate directly in the IGT, they participated directly in the USCMS Development Grid 
Testbed\footnote{The DGT is a smaller clone of the IGT on which it is permitted to deploy 
untested software.}(DGT) in support of the IGT efforts by troubleshooting problems and by 
doing important regression tests on the middleware.   

\begin{table*}[t]
\begin{center}
\begin{tabular}{||l|l|l|l|l||}
\hline IGT site & Worker CPU & CPU Speed & OS & Comments \\
\hline Caltech & 40 & 0.8 GHz & RedHat 6.X & \\
\hline Caltech & 40 & 2.4 GHz & RedHat 7.X & \\
\hline Fermilab & 80 & 0.75 GHz & RedHat 6.X & \\
\hline University of Florida & 80 & 1 GHz & RedHat 6.X & \\
\hline UC San Diego & 40 & 0.8 GHz & RedHat 6.X & Few CPUs inoperable after power failure\\
\hline UC San Diego & 40 & 2.4 GHz & RedHat 7.X & \\
\hline CERN & 72 & 2.4 GHz & RedHat 7.X & \\
\hline\hline UW Madison & 0 & N/A & Provided important SW support & \\
\hline
\end{tabular}
\caption{USCMS hardware currently dedicated to the Integration Grid Testbed.  Most
of this hardware will have been turned over onto a production grid by Spring 2003.}
\label{tb:hardware}
\end{center}
\end{table*}

In Spring 2003, most of this hardware will be turned over to a production grid.  
A small number of machines will be saved to do integration testing in support 
of production grid operations, which is the original purpose of the IGT in the 
first place.
 
\section{Software Running on the IGT}

  In the Grid environment, many different layers of software have 
to be present.  At the lowest level is the OS itself.  The IGT ran a Linux platform
with 
either RedHat 6.X based operating systems or with RedHat 7.X based 
operating systems\footnote{The use of RedHat 6.X based operating systems was 
required for CMS production with applications based on CMS ORCA which used
Objectivity as an object persistency layer.  This was not because of 
limitations within Objectivity, but was rather due to licensing issues.}. 
The Grid middleware was distributed using the Pacman \cite{bb:pacman} 
based Virtual Data Toolkit (VDT) \cite{bb:vdt}.  
The version which ran on the IGT was VDT 1.1.3.
The next level of software included the Job Creation level, consisting 
of mainly CMS developed tools and some PPDG provided tools to wrap 
CMS specific jobs in Grid aware wrappers, and finally the CMS applications 
themselves.

  Most Grid sites ran the PBS batch system or the Condor High Throughput Computing System, 
the Farm Batch System Next Generation (FBSNG) was run at Fermilab.\cite{bb:fbsng}
  
\subsection{VDT Middleware}

The VDT is composed of three types of grid software, although the
lines between these types are sometimes blurry. These three types are:
\begin{itemize}
\item \emph{Core Grid Software} This is the grid middleware,
and it includes Globus, \cite{bb:globus} Condor-G, 
and Condor \cite{bb:condor}. 
\item \emph{Virtual Data Software} This is the software that is able
to either compute or fetch data on demand, depending on whether it
needs to be computed or is already available.\cite{bb:virtualdata}
\item \emph{Utilities} This is a selection of software that provides
smaller but still important utilities, such as as software for
fetching certification authority revocation lists on a regular
schedule. 
\end{itemize}
The IGT made use primarily of the core Grid software and some of the 
utilities.

  The software in Table \ref{tb:software-vdt} was included with VDT version 1.1.3, 
which ran on the IGT.

\begin{table*}[t]
\begin{center}
\begin{tabular}{||l|l|l||}
\hline Component & Version & Comments \\
\hline Globus Toolkit & 2.0 & Modified Gatekeeper/Job Manager \\
\hline Condor & 6.4.3 & includes DAGMan \\
\hline Fault Tolerant Shell & 0.99 &  \\
\hline Globus Clients & 2.0 & eg- Globus-url-copy \\
\hline Condor-G & 6.4.3 & \\
\hline
\end{tabular}
\caption{Software from the VDT 1.1.3 currently installed on the IGT.}
\label{tb:software-vdt}
\end{center}
\end{table*}

\subsection{CMS Specific Software}

  The CMS software was distributed to the remote sites
using DAR \cite{bb:dar} before any jobs were submitted.
In principle, the CMS software can be installed as part of the job, as described 
below in the discussion of MOP.  However, the particular production request handled by 
the IGT was large enough (greater than 100,000 CPU hours on a single GHz CPU) to justify 
special pre-installation. 

  CMS Monte Carlo production consists of several steps \cite{bb:cms-prod-note}.  
First, vector representations
of simulated physics collision {\em events} are generated with the Pythia-based
CMKIN application. 
Second, the responses of the CMS detector are 
simulated in the GEANT 3 based 
CMSIM application.  A third CMS specific step is to re-format the 
CMSIM events into an Objectivity DB format with the writeHits application\footnote{This 
step will be combined with the previous step in future versions of the software.}.
Forth, the native detector signals must be mixed with noise simulations and with 
simulated by-products of nearly contemporaneous collisions called pileup (PU) in the 
writeDigis application.  Pileup events
are typically pre-processed and stored in locally resident files ready to be mixed with 
signal events in writeDigis, leading to I/O bound behavior when the number of pileup 
events per signal event is large.  (For the $10^{34}$ luminosity expected in the LHC, 
this ratio is about 200:1 on average.)  Finally, the last stage of production 
involves the creation of analysis objects (ntuples) to be analyzed by the physicists.
Table \ref{tb:CMS-exes} summarizes the average behaviors of the executables used in 
CMS Monte Carlo production.  Actual production depends most critically on the size of 
each event at the CMKIN stage: the more particles produced per event translated directly
into higher processing times.

\begin{table*}[t] 
\begin{center}
\begin{tabular}{|l|l|l|l|}\hline
Name&Time&Size\\
& (sec/event) & (MB/event/stage) & Bound\\ \hline\hline
CMKIN& 0.05 & 0.05 & \\ \hline
CMSIM & 350 & 2.0 & CPU Bound \\ \hline
writeHits & 0.05 & 1.0 & I/O Bound \\ \hline
writeDigis (NoPU)& 2.0 & 0.3 & CPU Bound \\ \hline
writeDigis (1034PU) & 10.0 & 3.0 & CPU and I/O Bound \\ \hline  
ntuple & $\le 1$ & 0.05 & CPU and I/O Bound \\ \hline  
\end{tabular}
\label{tb:CMS-exes}
\caption{CMS Executables with some running statistics on a 750 MHz machine. 
NOTE: The overall results can be highly variable depending on the physics 
being simulated, but the ratios should be about right.}
\end{center}
\end{table*}

  CMS production usually proceeds by breaking up production requests into sets of 
250 events each\footnote{250 events was found to be a reasonably large round number of events that lead to manageable filesizes and processing times.}
and processing each collection serially through all steps.  For the IGT production during
Fall 2002, there were two requests for events.  The first request was for 1M events
processed through all steps.  The second request was for 500K events processed only 
through the CMSIM stage.

\subsubsection{Job Creation: McRunjob}

 McRunjob \cite{bb:mcrunjob4} 
is a package containing Python scripts designed to aid in organizing
large scale CMS and DZero production processing applications.  
While initially developed specifically for DZero offline executables, 
McRunjob was easily extended to the CMS experiment.  
There are two basic classes within McRunjob.  The {\em Configurator}
encapsulates all of the knowledge needed to run an application or perform some 
simple task and exposes only the metadata with a customizable interface.  
The {\em Linker} is a container class for Configurators responsible for 
instantiation,  maintaining a list of Configurators, and handling communication 
among the Configurators and the user and other Configurators.  
The user interacts with the Linker, sending commands to attach and configure 
various Configurators.  Conceptually, McRunjob provides a language by which 
the use can specify a workflow pattern abstractly and then the McRunjob 
implementation takes care to turn it into a set of submittable jobs.
Depending on which modules are included, McRunjob can target Virtual Data
Language of the Chimera system or jobs using the SAM system at DZero. 
 For the IGT, McRunjob comes with a set of Configurators that target the
CMS legacy Impala runtime scripts which were used in official Spring 2002 
Monte Carlo production.  McRunjob produced 4000 jobs for the million event 
full production request, while the balance of 2000 CMSIM-only jobs were 
produced by the Impala scripts.  

\subsection{MOP}

  The jobs as produced by McRunjob and Impala for the IGT run were not specially
``grid aware.''  The connection to the grid was provided by a thin software
layer called MOP (Monte Carlo Production). MOP basically represents existing McRunjob
or Impala produced jobs as Directed Acyclic Graphs (DAGs).  There were four
generic types of DAG nodes that help accomplish this.  Stage-in nodes were responsible
for transporting the execution environment to the worker node.  Run nodes were 
responsible for running the executables.  Stage-out nodes were responsible for
transporting results back to the submit site.  Finally, clean-up nodes were 
responsible for removing any left over job state from the worker nodes.  

  During the IGT running, MOP was invoked to create DAG representations of each job
at job submit time.  Once a DAG was produced, MOP submitted the DAG to the DAGMan 
package of Condor.  DAGMan usually runs DAG nodes using the Condor system; for the IGT
Condor-G was used as a gateway to allow DAGMan to run DAG nodes on remote sites 
running Globus Job Managers.  In turn, these Globus Job Managers are able to run the
jobs using local batch queues.  For the IGT, only Condor and the Farm Batch System 
of Fermilab were used as local batch interfaces.

  Scheduling functionality was not implemented in the MOP system during the
Fall 2002 IGT run.  Rather, jobs were distributed by direct operator 
specification at job submission time.  MOP was logically divided into the MOP
master site and the MOP worker sites.  Jobs were created and submitted from the 
MOP master site, all input files were staged in from the MOP master site, and
all output was returned to the MOP master site.  No replica catalogs were used 
during the production process itself, but resulting data products were 
registered in a replica catalog 
at the end of processing. During Fall 2002, Fermilab hosted 
the IGT MOP master site, while UW Madison hosted a MOP master site for the DGT.

\subsection{Virtual Organization}

  The Virtual Organization (VO) was implemented using a system
to help us to organize the Grid mapfiles of Globus.  We used the 
Caltech Virtual Organization Group Manager \cite{bb:vo} developed
at Caltech to manage the VO users on the testbed. Group Manger 
stored the user information in an LDAP
database and organized the users into different groups. A
VO administrator could create/add groups and populate users. This
info could either be uploaded by 
user certificate or through the LDAP certificate server if 
the Department of Energy Science Grid (DOESG)
certificates were used. The VO tool was installed on a
server machine at Fermilab.  Each site (including Fermilab)
used the EDG mkgridmap script to generate the gridmap-file
entries based on the information stored in the LDAP database
at Fermilab. For the IGT Fall 2002 run, we used both
Globus certificates and DOESG certificates.

\subsection{Monitoring}

In the Fall 2002 IGT run, monitoring systems were not used for 
automatic controls such as scheduling, but rather for creating human 
readable displays only.  
Monitoring was divided logically into two different concerns.  
Local monitoring consists of gathering useful information from 
a single cluster or grid site, while Grid-wide monitoring consists 
of transporting and integrating information from local sites.  
IGT-wide monitoring was accomplished using the MonaLisa monitoring 
tool developed at Caltech \cite{bb:monalisa}.  Local monitoring was accomplished using 
modules written for MonaLisa at some sites and Ganglia at other sites.

\section{Running a Production Grid Continuously for the First Time}

The IGT ran CMS production by integrating CMS applications with Grid tools. 
Several problems were identified and fixed during this run. These include 
integration issues arising from non-grid CMS tools integrated with Grid 
tools, bottlenecks arising from operating system limitations (eg- default limits
not set high enough), and bugs in 
middleware and application software. Every component of the software 
contributed to overall problem count in some way. However, we found that 
with the current level of functionality, we were able to operate the 
IGT with 1.0 FTE effort during quiescent times over and above normal 
system administration and up to 2.5 FTE during crises.  A sampler of 
problems appears below.

\begin{itemize}
\item (Pre-IGT) During Spring 2002, the Globus 2.0 GASS Cache was found to not support the 
required level of performance for CMS production. The software was re-engineered
in consultation 
with Condor developers and Globus developers over the summer of 2002, and released 
in Globus 2.2.  
\item The Impala environment requires many helper files to run a production job.
It was found that many simultaneous globus-url-copy operations originating from 
the MOP master site when submitting many jobs would cause some globus-url-copy 
operations to hang.  Globus-url-copy operations were wrapped in Fault Tolerant Shell 
(FTSH) scripts.  FTSH contains semantics to time-out and retry shell commands.
\item  Keeping in view the need of automatic resubmission of failed jobs; 
MOP used the auto restart features of Condor\_G.  There have been several cases 
of this when actual jobs were still running on remote site.  While the jobs 
eventually finish, there was wasted CPU time.
\item  There were some instances when jobs failed due to application code problems, or 
some real reason like full disk partitions. With auto-restart option, it 
becomes difficult to find out actual cause of trouble, while failed jobs are 
restarted and resubmitted in an infinite loop. The actual problem remain 
hidden unless jobs are watched closely. 
\item Condor\_G running on the MOP Master site uses 'gahp\_server' to handle its 
communication with processes running under Globus on remote worker sites, 
one thread per tracked process. 
With over 400 CPUs available to IGT at later stages of production, running 
two assignments to produce 1.5 Million events, we had to divide production 
over two physically separate MOP master machines, to avoid scaling limit
of the number of gahp\_server threads.  
\end{itemize}

\section{Analysis Environment on the IGT: CLARENS}

A need was identified to provide a simple universal access method or
"portal" to data and CPU resources as part of the end-user physics analysis
process. Clarens \cite{bb:clarens} is a flexible web services layer 
accessible through SOAP or
XML-RPC which is designed for high security, high throughput request
processing. Server functionality is easily extensible through administrator-
or user-installed server-side modules written in C/C++ or the Python
scripting language. Individual service requests are handled by a
multi-process server, providing crash protection and support for
non-blocking long-lived requests.
  
Security is based on X509 certificates for authentication, and optional
transport encryption using SSL/TLS. A full Virtual Organization (VO)
implementation coupled with fine-grained access control lists (ACL) provide
powerful and easy to administer security for server methods and files. The
traditional mode of certificate to user mapping for execution of server-side
jobs is also supported.
  
Server methods available include file repository access, VO/ACL
administration, job execution, and a remote interface to the Sql2ROOT/SOCATS
(Stl based Object Caching And Transport System) RDBMS-based analysis
framework. Client access to server methods is through ROOT (command line,
interpreted and compiled C++), Python (command line and script-based), and
Java.

\section{SC2002 Demo}

A continuously updated display of simulated event production being run on
the IGT as part of the Fall 2002 production was demonstrated at the
Fermilab/SLAC booth at Supercomputing 2002 (SC2002) \cite{bb:sc2002}. 
The display was constructed using the ROOT analysis
package using data obtained from Clarens servers involved in the IGT
production at Caltech, Fermilab, UCSD and UFL. The data used were ROOT files
produced in the last step in the production process at each site. The
Clarens ROOT client continuously monitored the servers for newly produced
data files, which were incorporated into the analysis on the show floor. 
  
A second demo at the Caltech Center for Advanced Computing Research (CACR)
booth showed an interactive analysis of event data stored at many remote RDBMS
analyzed via the Clarens interface to SOCATS. The results of the analyses
performed at Caltech and the Startlight point of presence (POP) 
in Chicago were made available
as ROOT files that were interactively accessed through the ROOT Clarens
client.

\section{Conclusions}

\begin{figure*}[t]
\centering
\includegraphics[width=135mm]{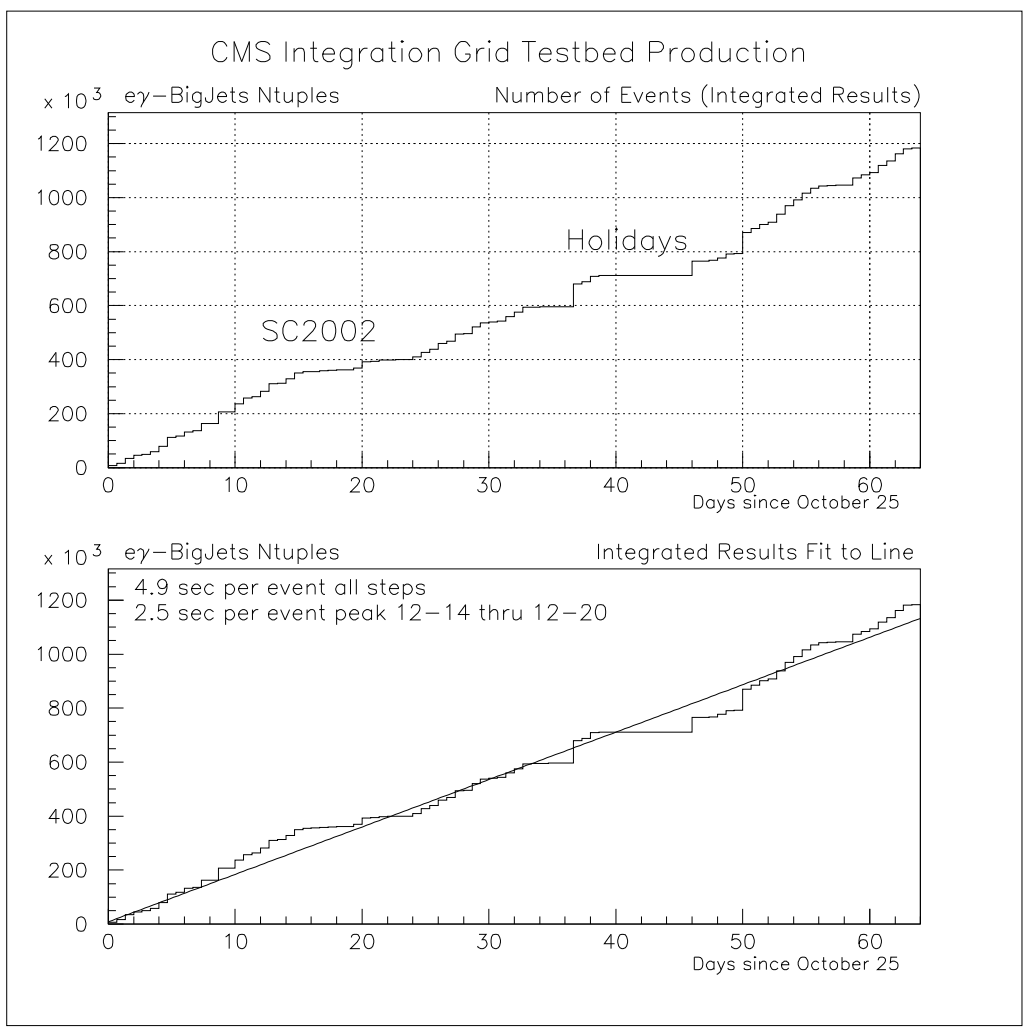}
\caption{IGT production progress during Fall 2002.}
\label{fg:igt-1}
\end{figure*}

The IGT was a success in that it produced all of the required events 
and provided many useful insights into operating a grid in production 
mode.  Also, many problems were uncovered with the software at all 
levels.  Figure \ref{fg:igt-1} shows the progress of IGT full ntuple 
production during Fall of 2002.

Despite the problems, the production was remarkably smooth and sustained
for over two months.  The two notable flat spots occur during the SC2002 
conference and during the winter holidays, which reflect loss of manpower
during those periods.  

  In order to better quantify efficiency, the IGT run period was divided into 
12 periods of about five days each. The average daily production rate in each 
interval was compared to the theoretical maximum daily rate of 45 K events 
per day IGT-wide\footnote{This number was estimated by scaling the daily 
throughput observed on one machine to all machines weighted by rated CPU speeds.}.  
The average efficiency was just under 40\%.

\begin{figure*}[t]
\centering
\includegraphics[width=135mm]{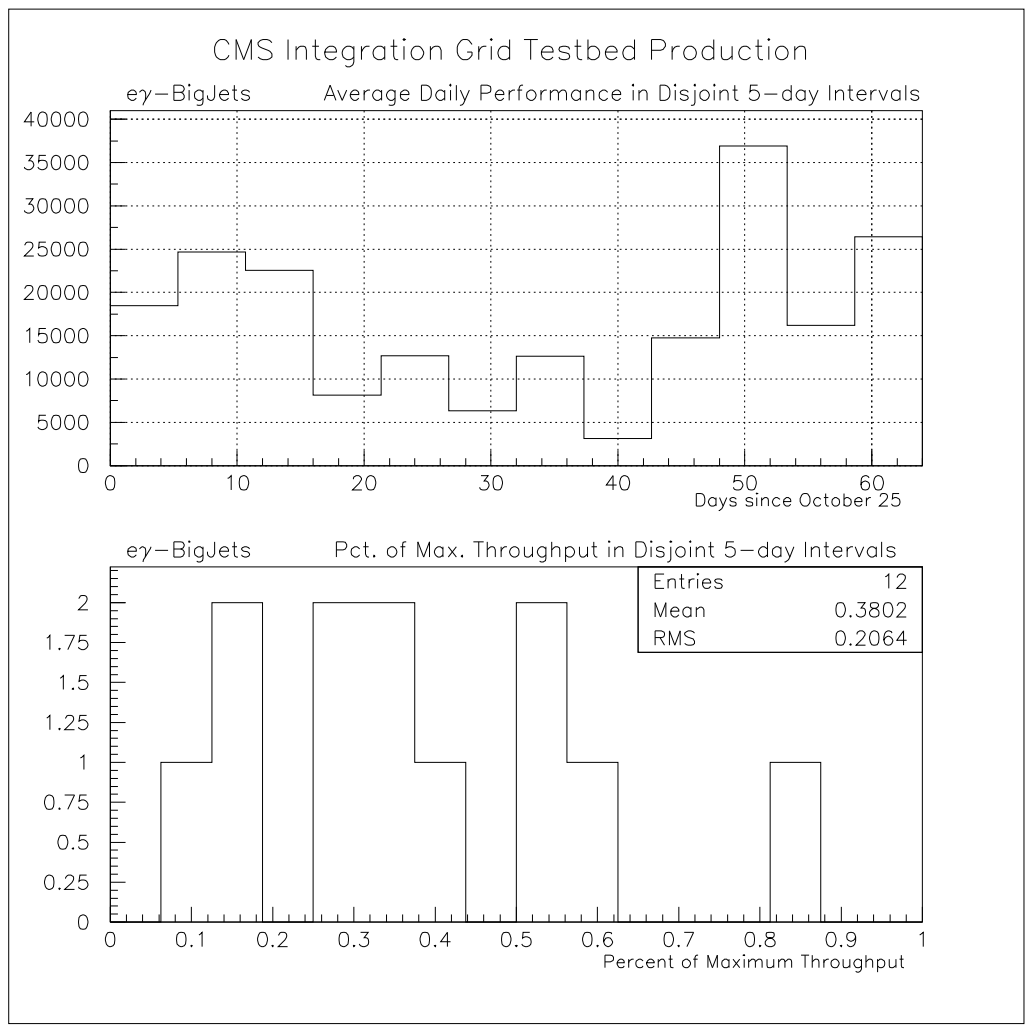}
\caption{Efficiency of the IGT during Fall 2002 running.  Based on the 
amount of resources available and the measured program performance on a 
single CPU, a theoretical maximum throughput of 45K events per day was calculated.
Stated efficiencies are relative to that maximum throughput.}
\label{fg:igt-2}
\end{figure*}

  This performance is not much worse than the conventional 
CMS Spring 2002 Production.  The Spring 2002 production was more complicated 
in that it involved more events with pileup and involved a lot more file 
transfers.  Also, it is hard to calculate efficiency of the Spring 2002 production
because it is hard to determine when a site was unavailable due to problems or just idle 
for lack of a request.   

  The EDG stresstest \cite{bb:edg-stresstest} ran during Fall of 2002 also.  
The EDG stresstest involved 
more functionality than the IGT in that it used a resource broker which relied
on MDS to supply it with timely information.  In short, they found that with the current 
state of Grid middleware, more functionality led to more problems,
(and more FTE expended to track them down.)  Viewed for what it was, the EDG stresstest
was an immense success and found problems that were complementary to those
found in the contemporaneous IGT run.   Finally, our experience with the IGT 
has led to a plethora of documentation 
\cite{bb:igt1},\cite{bb:igt2} that can be a good start to providing production level 
support.

\begin{acknowledgments}
We would like to acknowledge the CMS Core Computing and Software 
group and the USCMS Software and Computing projects for supporting this 
effort.  We would especially like to thank Veronique Lefebure and Tony
Wildish of the CMS Production Team for their support and helpful 
discussions. 
This effort was also supported by the USCMS Development Grid Testbed,
including additional testbed hardware and support from the University of Wisconsin at 
Madison.
\end{acknowledgments}


\end{document}